\documentstyle[psfig,conf_iap]{article}
\begin{document}
\vglue-3cm
{\small \sl \noindent Contributed talk at XIVth IAP meeting ``Wide-Field
Surveys in Cosmology'', Paris, May 1998,
ed.  Y. Mellier \& S. Colombi (Paris: Fronti\`eres)}
\vspace{1.5cm}

\heading{The Wide-Field DENIS Near-IR Imaging Survey and 6dF
Redshift and Peculiar Velocity Surveys} 
\par\medskip\noindent
\author{Gary A. Mamon$^{1,2}$}
\address{Institut d'Astrophysique de Paris, 98 bis Boulevard Arago,
F--75014 Paris, France}
\address{DAEC, Observatoire de Paris, F--92195 Meudon, France}

\begin{abstract}
The {\sf DENIS} survey is currently imaging the entire southern sky in the
$I$, $J$, and $K$ wavebands. The current star/galaxy separation algorithm is
presented and the galaxy counts are nearly perfectly Euclidean.  95\%
complete and reliable galaxy samples with better than 0.2 magnitude
photometry should include $50\,000$ ($K < 12.0$), $500\,000$ ($J < 14.8$), and
$900\,000$ ($I < 16.5$) galaxies, respectively, over the full hemisphere.
Two spectroscopic followups of {\sf DENIS} and {\sf 2MASS} galaxies are
planned on the {\sf 6dF} robotic multi-object spectroscopic unit, currently
under construction at the {\sf AAO}, and which will be mounted on the {\sf
UKST} Schmidt telescope: a redshift survey of roughly $120\,000$ NIR selected
galaxies and a peculiar velocity survey of roughly $15\,000$ galaxies (both
early-types and inclined spirals) at $cz < 10\,000 \, \rm km \, s^{-1}$.
\end{abstract}
\section{Importance of Near-IR selected nearby galaxy samples}

The Near-IR (NIR) wavebands are considered to provide galaxy catalogs
that  optimally trace the  distribution of matter in the
Universe.
Indeed, cosmological studies have all been based upon optical
({\it e.g.\/,} \cite{maddox}) or far-IR ({\it e.g.\/,} \cite{qdot})
surveys, 
and the optical wavebands are severely affected by extinction from
interstellar dust, while both optical and far-IR wavebands tend to pick up
galaxies undergoing recent starbursts.
For this reason, the European {\sf DENIS} $IJK$ survey of the southern
hemisphere \cite{denis} and the American {\sf 2MASS} $JHK$
survey of the entire sky \cite{2mass} have important cosmological
implications \cite{mamon_moriond96}.
In essence, all the cosmological analyses of the distribution of
matter in the local Universe, performed in the optical or the far-IR, need to
be redone in the NIR.

\section{DENIS galaxy extraction}

The current {\sf DENIS} galaxy pipeline is based upon fairly standard
procedures, described elsewhere \cite{mamon_EC3}\cite{mamon_AA}.
Star/galaxy separation is performed in the $I$
band, even for $J$ and $K$-selected objects, because the $I$-band is the most
sensitive (except at $|b| < 3^\circ$) and has the 
best spatial resolution. 

Excellent star/galaxy separation is crucial,
because, at the fairly bright magnitude limits of {\sf DENIS}, stars
vastly outnumber galaxies, and also because the point spread function (PSF)
of the {\sf DENIS} images varies in space and in time.  We have studied
various star/galaxy separation algorithms, by comparing their output to
visual classification of 329 galaxy candidates and to the output of the {\sf
APM} \cite{apm_cat} and {\sf EDGSC} \cite{cosmos_cat} galaxy catalogs.  This
led 
us to abandon neural networks, which were improperly trained for PSF
variations in space and time.  The ratio of peak intensity to isophotal area
(hereafter RATIO) turns out to be perhaps the simple star/galaxy separation
algorithm that is the most robust to variations of the
PSF \cite{mamon_EC3}\cite{mamon_AA}.

%
\vspace{-2mm}
\begin{figure}[h]
\centerline{\vbox{\psfig{figure=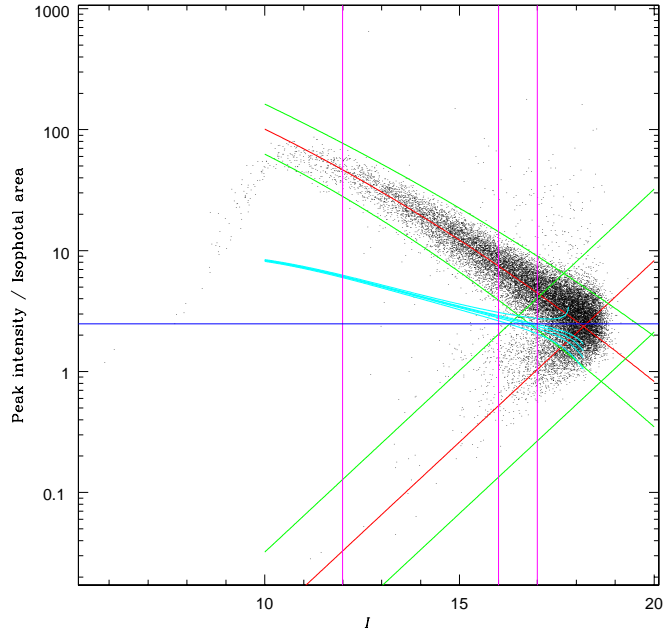,height=9.2cm}}}
\caption[]{
Star/galaxy separation plot for 180 consecutive {\sf DENIS} $I$-band images
at high galactic latitude.
The {\it upper locus\/} shows the foreground stars, and shows saturation at
$I \leq 
11$.
The {\it lower locus\/} shows the galaxies.
The {\it curved\/} and {\it straight lines\/} around the star and galaxy
locii respectively 
represent our  
estimates of  their
mean and 3$\sigma$ envelopes.
The {\it horizontal line\/} shows a 1st-order star/galaxy separation, which
works 
well for $I \leq 16.5$.
The family of {\it nearly horizontal curves\/} represent our 2nd-order
star/galaxy 
separation, as described in the text, with reliabilities of 50\%, 84\%, 90\%,
95\% and 99\% (see eq. [\ref{eq:relia}]) from top to bottom.
The {\it vertical lines\/} show the magnitude intervals where the stellar
locus is 
defined ({\it outer two lines\/}) and where the star counts are estimated
({\it left two lines\/}).
Note that this star/galaxy separation is not meant to work beyond where our
estimates of the star and galaxy locii cross.
}
\end{figure}
Figure~1 above shows our star/galaxy separation on a sequence of 180
consecutive {\sf DENIS} $I$-band images at high galactic latitude.
A constant RATIO threshold works well for
$I \leq 16.5$, but poorly at fainter magnitudes.
A better star/galaxy separation is obtained by computing critical curves for
\begin{equation}
R = {f_g(m) N(x,\bar x_g(m), \sigma_g(m))
\over
f_g(m) N(x,\bar x_g(m), \sigma_g(m)) +
f_s(m) N(x,\bar x_s(m), \sigma_s(m))}
\ ,
\label{eq:relia}
\end{equation}
where
$N(x,\mu,\sigma)$ is the gaussian of mean $\mu$ and standard deviation
$\sigma$, $f$ are the number counts, $x$ is our star/galaxy separation
statistic, and subscripts `$g$' and `$s$' refer to galaxies and stars,
respectively.
%
%
%
We've tested the reliability of our two methods, by comparing
with visual classifications from {\sf DENIS} images and with {\sf APM} and
{\sf COSMOS} classifications, and by checking the very low frequency of very
blue or red colors for our galaxies \cite{mamon_EC3}\cite{mamon_AA} and we
have begun to test the spatial uniformity along the image frames of our
extraction.

\begin{figure}[h]
\centerline{\psfig{figure=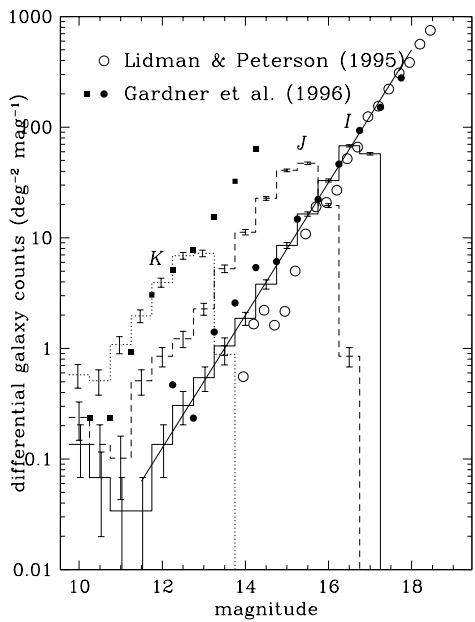,height=8.7cm,width=6.5cm}}
\caption[]{
$I$ ({\it solid\/}), $J$ ({\it dashed\/}) and $K$ ({\it dotted histogram\/})
galaxy 
counts from  
$50 \,\rm deg^2$ of high galactic latitude  {\sf DENIS} images.
The {\it solid line\/} is an eye-ball fit of the $I$ counts to a Euclidean
(0.6) 
slope.
}
\end{figure}
Comparing our resulting galaxy
counts in Figure~2 to the previous published counts, we infer $>90\%$
completeness
limits of $I \simeq 16.5$, $J \simeq 14.8$, and $K \simeq 12.0$.
These yield homogeneous samples of 
$900\,000 \,(I)$,
$500\,000 \,(J)$,
and
$50\,000 \,(K)$ galaxies over the southern sky.
The nine-fold increase of our estimated $K$ sample, relative to our
previous estimates ({\it e.g.\/,} \cite{mamon_EC3}), is the consequence of the
much wider smoothing filter we now use to detect galaxies in the $K$ band.
Our counts show Euclidean slopes and no lack of bright objects, in
contrast with the {\sf APM} \cite{apm_counts} and {\sf EDGSC}
\cite{cosmos_counts} counts. 

\section{Spectroscopic followups with 6dF}

Simulations \cite{mamon_moriond96} of the duration time of a NIR-selected full
redshift survey show that a wide field telescope is necessary
to tile the sky efficiently.
Based upon these findings, we've been able to convince the {\sf AAO} to
robotize its currently highly inefficient {\sf FLAIR-II} multi-object
spectrograph on the {\sf UKST}, increasing from 92 to 150 its number of fibers.
The new instrument, currently under construction, is called {\sf 6dF} for
{\sf Six Degree Field}.
The {\sf AAO} 
is suggesting dedicating 100 nights per year during 2001--2003 to obtain
roughly 
$120\,000$ redshifts.
Our input catalog will consist of a combination of magnitude-limited
subcatalogs from  {\sf DENIS} $IJK$, {\sf
2MASS} \cite{2mass} $JHK$, and hopefully {\sf
SUPERCOS} \cite{supercos} $B$ 
and $R$.


We are also planning a peculiar velocity survey of roughly $15\,000$
galaxies. The sample will be NIR flux-limited, and restricted to distances
$cz < 10\,000 \, \rm km \, s^{-1}$ (determined from the redshift survey).
We will measure the distances to both early-types and inclined spirals and
infer the peculiar velocities by subtracting the Hubble expansion velocity from
the radial velocity.

Our main scientific aim is to map the mass density field of the local
southern Universe and to infer $\Omega$, $\Lambda$ and the
primordial density fluctuation spectrum.
For the first time, 
we will provide complete and uniform coverage, multiplying by 10 the current
sample of peculiar velocities in the southern
hemisphere.

%
\vfill

\begin{iapbib}{99}{
\bibitem{denis} Epchtein, N. et al. 1997, Messenger 87, 27
\bibitem{gardner} Gardner, J.P.,
 Sharples, R.M.,
 Carrasco, B. \& Frenk, C. 1996, MNRAS  282, L1
\bibitem{supercos} Hambly, N.C., Miller, L., MacGillivray, H.T.,
Herd, J.T. \& Cormack, W.A. 1998, MNRAS 298, 897
\bibitem{cosmos_counts} Heydon-Dumbleton, N.H., Collins, C.A. \&
MacGillivray, H.T. 1989, MNRAS 238, 379
\bibitem{2mass} Kleinmann, S.G. et al. 1994, Ap\&SS 217, 11
\bibitem{lidman} Lidman, C.E. \& Peterson, B.A. 1996, MNRAS 279, 1357
\bibitem{maddox} Maddox, S.J., Efstathiou, G., Sutherland, W.J. \& 
Loveday, J. 1990, MNRAS 242, 43P
\bibitem{apm_counts} Maddox, S.J., Sutherland, W.J., Efstathiou, G.,
Loveday, J. \&  Peterson, B.A. 1990, MNRAS 247, 1P
\bibitem{mamon_moriond96} Mamon, G.A. 1996, ed Ansari, R. et al., in
{\it Dark matter in cosmology, quantum measurements, and experimental
gravitation\/}, p. 225 (astro-ph/9608076)
\bibitem{mamon_EC3} Mamon, G.A., Borsenberger, J., Tricottet, M. \& Banchet,
V. 1998, ed Epchtein, N., in {\it The impact of near-infrared sky surveys on
galactic and extragalactic astronomy\/}. Kluwer, Dordrecht, p. 177
(astro-ph/9712169)
\bibitem{mamon_AA} Mamon, G.A., Bertin, E., Borsenberger, J., Epchtein, N.,
Fouqu\'e, P. \& Tricottet, M. 1998,  to be submitted to A\&A
\bibitem{qdot} Saunders, W., Frenk, C.,
Rowan-Robinson, M., Lawrence, A. \& Efstathiou, G. 1991, Nature 349, 42 
\bibitem{apm_cat} {\tt telnet://catalogues@apm3.ast.cam.ac.uk}
\bibitem{cosmos_cat} {\tt telnet://cosmos@cosmos.aao.gov.edu.au}
}
\end{iapbib}
\end{document}